\documentclass[prl, preprint, superscriptaddress, showkeys]{revtex4-2}

\usepackage{hyperref}
\hypersetup{colorlinks=true, citecolor=blue, urlcolor=blue, linkcolor=blue}

\usepackage{lineno}
\usepackage{graphicx}
\usepackage{amsmath}
\usepackage{cases}

 \newcommand{\mpq}{Laboratoire Matériaux et Phénomènes Quantiques (MPQ), Université Paris Cité, 75205 Paris, France}
 \newcommand{\ctwon}{Centre de Nanosciences et de Nanotechnologies, CNRS, 91120 Palaiseau, France}
 \newcommand{\diet}{DIET, Sapienza University of Rome, Via Eudossiana, 00184 Roma, Italy}
 \newcommand{\cnr}{CNR-INO, Istituto Nazionale di Ottica, Via Campi Flegrei 34, 80078 Pozzuoli, Italy}
 \newcommand{\Lu}{Department of Physics, Chemistry and Biology, Linköping University, Linköping, SE-581 83, Sweden}
 \newcommand{\ulb}{Service OPERA-Photonique, Université libre de Bruxelles (ULB), B-1050 Brussels, Belgium}
 \newcommand{\KIT}{Institute of Photonics and Quantum Electronics (IPQ), Karlsruhe Institute of Technology (KIT), 76131 Karlsruhe, Germany}
 
 \newcommand{\ual}{Applied Physics, Department of Chemistry and Physics, University of Almeria, 04120 Almeria, Spain}

\begin{document}

\title{Bistable soliton optical frequency combs in a second harmonic generation Kerr cavity}

\author{Francesco Rinaldo Talenti}
\affiliation{\ctwon}
\affiliation{\mpq}

\author{Stefan Wabnitz}
\affiliation{\diet}
\affiliation{\cnr}
\author{Yifan Sun}
\affiliation{\ulb}

\author{Tobias Hansson}
\affiliation{\Lu}

\author{Luca Lovisolo}
\affiliation{\mpq }
\affiliation{\ctwon }

\author{Andrea Gerini}
\affiliation{\mpq }

\author{Giuseppe Leo}
\affiliation{\mpq}

\author{Laurent Vivien}
\affiliation{\ctwon}

\author{Christian Koos}
\affiliation{\KIT}

\author{Huanfa Peng}
\affiliation{\KIT}

\author{Pedro Parra-Rivas}
\affiliation{\ual}

% \affiliation[*]{francesco-rinaldo.talenti@universite-paris-saclay.fr}
% \affil[*]{pedro.parra-rivas@ual.es}

\begin{abstract}
We study the dynamics and stability of soliton optical frequency comb generation in a dissipative, coherently pumped cavity with both second and third-order nonlinearity. Cavity sweep simulations and linear stability analysis based on path continuation reveal the existence of bistable solitons. These families of solutions represent a continuous transition between a purely quadratic and a Kerr cavity soliton frequency comb. Perspective demonstrations of these novel optical sources is an ongoing relevant subject within the frequency comb community.
\end{abstract}

\maketitle

% \section{Introduction}

The propagation of optical pulses as cavity solitons (CSs) is self-sustained by the interplay between dispersive and nonlinear effects in dissipative and coherently pumped cavities \cite{Wabnitz:1993}. Their spectrum forms a broad and coherent optical frequency comb (OFC). The experimental generation of soliton microcombs \cite{Del'Haye:2011} is based on triggering modulation instabilities \cite{Haelterman:1992}, which leads to either chaotic, periodic, or stationary CS formation. Despite the robustness of their dynamical basin of attraction, the chaotic origin of CSs leads to the intrinsic difficulty of deterministically exciting a specific soliton state \cite{Rowley:2022}. For this reason, different schemes for the control and stabilization of CSs have been proposed and demonstrated \cite{Rowley:2022,Talenti:2023}.\\ 
An interesting class of OFC is based on purely quadratic nonlinear cavities (i.e. $\chi^{(2)}\ne 0$, $\chi^{(3)}\sim 0$), when coupled combs are generated around the fundamental frequency (FF) and the second-harmonic (SH) \cite{Ricciardi:2015}.  In this case, it has been shown that the resulting dynamics for the FF can mimic an effective Kerr nonlinearity, leading to different OFC regimes \cite{Leo:2016_PRL}. This analogy is particularly relevant when the cavity is only resonant for the FF, which permits to derive an effective nonlinear Schrödinger equation (NLSE) with coherent driving and damping. For a doubly resonant cavity, a pure quadratic OFCs can be modeled by means of two coupled nonlinear mean-field equations \cite{Leo:2016_PRA,Hansson:2018}. A double (or multi) envelope approach can be also adopted for Kerr OFCs  (i.e. $\chi^{(2)}\sim 0$, $\chi^{(3)}\ne 0$) \cite{Skryabin:1999, Hansson:2023}, under the general assumption that the resonant field owns components in separate spectral domains. 

In waveguides with both quadratic and cubic nonlinearities (i.e. $\chi^{(2)}\ne 0$, $\chi^{(3)}\ne 0$), self- (SPM) and cross- (XPM) phase modulation can trigger modulation instabilities \cite{Trillo:1992} and soliton formation \cite{buryak:1995}. In these systems, different families of two-wave solitons can be generated. A characteristic CS solution is formed comprising an intense and a weak pulse component around the FF and the SH, respectively. The opposite case, where most of the energy is carried by the SH, has also been theoretically studied \cite{Trillo:1992}. A bistable regime may arise, whenever the two families of CSs coexist in the same point of a multi-dimensional phase diagram, spanned by some characteristic dynamical parameters. as recently predicted for third-harmonic-generation cavities \cite{Hansson:2023}. %Under specific conditions bistable solitons are also found in single envelope mean field NLSE models, whereas for typical Kerr or Pockels systems one typically obtains singly-valued CS solutions \cite{kaplan:bistableCS}. 
As we shall see in this work, the doubly resonant condition is of crucial importance to observe bistable CSs in $\chi^{(2)}$ and $\chi^{(3)}$ resonators. While an optical parametric oscillator (OPO) is discussed in Ref.\cite{Villois:2019}, here we study the case of a second harmonic generation (SHG) cavity. This involves a system of two coupled equations including SHG, SPM and XPM terms \cite{Xue:2017}, whose solutions feature a bistable CS regime. Optical bistability results from the Kerr effect acting on both FF and SH via SPM and XPM. Moreover, CS bistability is observed in situations where a significant fraction of the FF is coupled to the SH.\\
%This responds to the necessary condition of having an important harmonic contribution to the nonlinear dynamics. 
%The bistable behavior, however, rely on SPM and XPM terms. 
%\subsection*{The model}
\begin{figure}[ht!]
\centering
\includegraphics[width=0.5\linewidth]{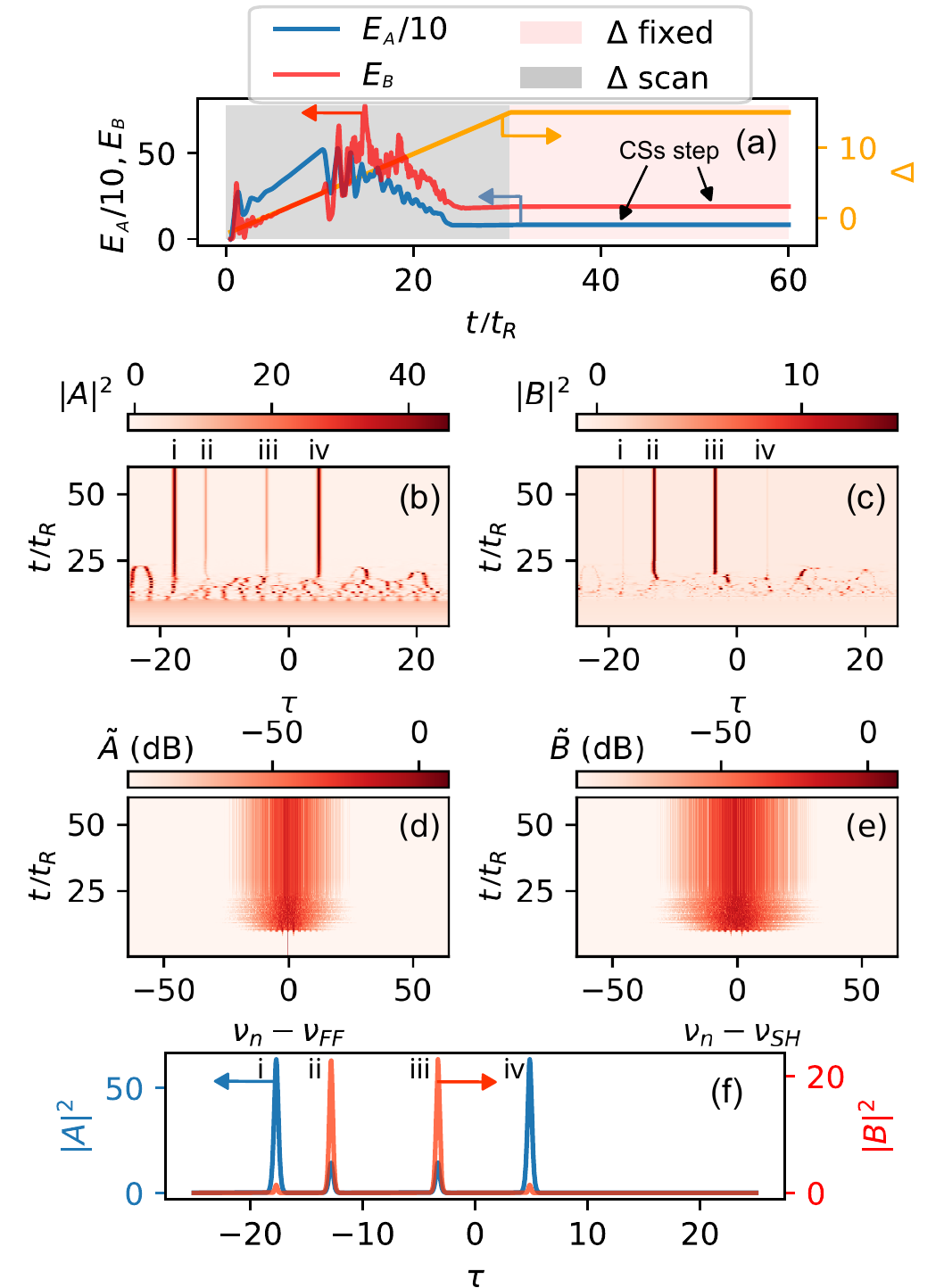}
\caption{(a) Slow-time evolution of cavity detuning $\Delta$ (orange) and intra-cavity energy of FF $E_A$ (blue) and SH $E_B$ (red) fields, respectively. Temporal (b-c) and spectral (d-e) evolution of intra-cavity $A$ and $B$ intensities. (f) Final bistable CSs state. In panels (a) and (f) the colored arrows indicate the y-axis for the corresponding quantity.}
\label{fig:DNS_pcolor}
\end{figure}
%\begin{figure}
%\includegraphics[width=0.9\linewidth]{FIGS/DNS_FIG2.png}
%\caption{Final state of the cavity sweep performed. In (a) we report the temporal intra-cavity field, while in (b-c) the spectrum of the FF and SH field.}
%\label{fig:DNS_figs}
%\end{figure}

Let us adopt a double-envelope approach \cite{Phillips:24} which considers the set of coupled, coherently driven and damped NLSEs with the contribution of SHG terms \cite{Leo:2016_PRA,Hansson:2018,Xue:2017}:

\begin{align}
  &  \dfrac{\partial A}{\partial t}=\left[ -\alpha_1-i \Delta-i\eta_1 \dfrac{\partial^2}{\partial \tau^2} \right] A +i \kappa B A^* + S  \nonumber %\label{eq:L1}
   \\
   & \ \ \ \ \ \ \ +\left[ i \gamma_1 |A|^2 + 2 i \gamma_{12} |B|^2 \right] A  \label{eq:NL1} \\
   & \dfrac{\partial B}{\partial t} =  \left[ -\alpha_2 -i2 \Delta  - d\dfrac{\partial}{\partial \tau}  -i\eta_2 \dfrac{\partial^2}{\partial \tau^2}  \right] B+i\kappa A^2 \nonumber %\label{eq:L2} 
   \\
   & \ \ \ \ \ \ \ + \left[i \gamma_2 |B|^2 + i2 \gamma_{21} |A|^2\right] B \ \ \ \ \ ,\label{eq:NL2}  
\end{align}
where $A$ and $B$ are the wave amplitudes at FF (index 1) and SH (index 2), respectively. The linear response of the cavity 
%(eqs.(\ref{eq:L1},\ref{eq:L2})), 
driven by the external laser source $S$ includes distributed losses ($\alpha_{1,2}$), laser/cavity detuning ($\Delta$), walk-off ($d$) and group velocity dispersions ($\eta_{1,2}$); SHG ($\kappa$), SPM ($\gamma_{1,2}$) and XPM ($\gamma_{12,21}$) terms provide the nonlinear response of the cavity, while we neglect higher-order dispersive terms and suppose perfect phase matching. 
%(eqs.(\ref{eq:NL1},\ref{eq:NL2})). 
The FF and SH intra-cavity fields evolve in two separate time-scales \cite{Hansson:2018}, the \textit{slow} ($t$) and the \textit{fast} ($\tau$) time, respectively. The former describes the evolution of the fields over successive round trips, while the latter measures the time dependence of the intra-cavity field. Generally, to generate a broadband OFC in $\chi^{(2)}+\chi^{(3)}$ ring cavities, besides the anomalous dispersion condition, both phase and group-velocity matching between the FF and SH waves are required  \cite{Talenti:2024}. For this reason, we consider here a zero group-velocity walk-off regime ($d=0$); however, direct numerical simulations (DNS, not reported here) show that stable CSs also survive in the presence of nonzero walk-off. 

To show that two-wave stable  $\chi^{(2)}+\chi^{(3)}$ CSs can be effectively generated, let us present a DNS of Eqs. (\ref{eq:NL1}-\ref{eq:NL2}) with the parameter values: $S=6$, $\alpha_{1,2}=1$, $\eta_{1,2}=-1$, $\gamma_{1,12}=0.5$, $\gamma_{2,21}=1$, and $\kappa=1$. To check the stability of the excited dynamical states, let us consider a linear ramp of cavity detuning $\Delta$ over the first 30 round-trips (from $\Delta=-2$ up to $\Delta=15$), and then keep it constant. Fig. \ref{fig:DNS_pcolor} (a) shows the build-up of the related intra-cavity energies $E_{A,B}(t)$ at FF and SH, which are computed as the fast time integrals of the respective intra-cavity optical intensities $|A|^2$, $|B|^2$. The emergence of the CSs step is clearly visible for both the FF and SH fields, and its associated fixed value, once that $\Delta$ remains a constant, is a direct indication that a stable physical state has been excited. In panels (b-c) and (d-e) of Fig. \ref{fig:DNS_pcolor} we show the temporal and the spectral DNS evolution of the FF and SH fields, respectively. The spectrum ($\tilde{A}$, $\tilde{B}$) is normalized to its maximum, and reported in a dB scale. 

From the time-domain plots, we may note the emergence of four CSs, which remain stable once that the detuning is fixed [these CSs marked in Roman numbers i-iv in panels (b,c)]. Interestingly, we observe that the CSs have different energies. For two of them (i,iv), most of the power is carried by the FF wave, while the other two (ii,iii) have a dominating SH component. This is a clear signature of solitons bistability, similar to what previously observed in the case of third-harmonic generation \cite{Hansson:2023}. To highlight this property, we show the final state of the simulation in Fig.\ref{fig:DNS_pcolor} (f) . As we can see, one may clearly regroup the CSs into two families: for the CS (i,iv), the FF dynamics dominates, while for the solitons in (ii,iii) the SH contribution is largely dominant. Similar solutions were previously found in a Kerr guided system \cite{buryak:1995}. It is worth to mention that the CSs i-iv are sustained by a nonzero homogeneous background, which is not fully appreciable in panel (f) owing to the large CS peaks to background aspect ratio.

%\subsection*{Bifurcations}
A highly non-linear system generally has a plethora of possible solutions. As a rule of thumb, the stronger the nonlinear effects, the richer the dynamics. In practice, system (\ref{eq:NL1}-\ref{eq:NL2}) easily turns out to be very chaotic, and it is hard to locate stable solutions in the phase space spanned by the key dynamical parameters, such as the driving field $S$ and the cavity detuning $\Delta$. To systematically quantify the contribution of the different nonlinear terms, one may rewrite Eqs.(\ref{eq:NL1})-(\ref{eq:NL2}) in operator form as follows:
\begin{equation}
  \dfrac{\partial \mathcal{U}}{\partial t}=\mathcal{F}  \equiv\mathcal{L}\mathcal{U} + \mathcal{Q}\mathcal{U} +\sigma \mathcal{K}\mathcal{U}+\mathcal{S}  \ \ \ ,
    \label{eq:system}
\end{equation}
\noindent where $\mathcal{U} =(A, B)$, $\mathcal{L}$, $\mathcal{Q}$, $\mathcal{K}$ are the linear, quadratic, and cubic operators, and $\mathcal{S}$ is the external driving term, implicitly defined by comparison with Eqs.~(\ref{eq:NL1})-(\ref{eq:NL2}). We have also introduced the dimensionless parameter $\sigma\equiv\gamma_{2}/\kappa$, which controls the nonlinear interplay and quantifies the contribution of the Kerr operator $\mathcal{K}$. This definition is subject to a rescaling of the $\gamma$ factors: $\gamma_{ij}'=(\gamma_{ij}\kappa)/\gamma_2$, with $ij=1,2,12,21$. 

%source  are defined consistently with Eqs.(\ref{eq:NL1})-(\ref{eq:NL2}) with $d=0$:

%----------------------------------- 

%\begin{align}
 %   \mathcal{L}\mathcal{U}= \begin{pmatrix}
%-\alpha_1-i \Delta +i \eta_1\partial^2_{\tau}\\
%-\alpha_2-i 2 \Delta +i \eta_2\partial^2_{\tau}
%-\alpha_2-i 2 \Delta +i \eta_2\partial^2_{\tau}
%\end{pmatrix}& \mathcal{U}
   % \label{eq:Loperator}
 %   \qquad
 %  \mathcal{S}= \begin{pmatrix}
%S\\
%0
%\end{pmatrix} 
    %\\
    %\mathcal{Q}\mathcal{U}=i\begin{pmatrix}
%\kappa B A^*\\
%\kappa^* A^2
%\end{pmatrix}, \ \ \ %\mathcal{K}\mathcal{U}=&\begin{pmatrix}
%i \gamma_1|A|^2+ 2 i \gamma_{12}|B|^2\\
%i \gamma_2|B|^2+ 2 i \gamma_{21}|A|^2
%\end{pmatrix} \mathcal{U} \ \ \ .
 %   \label{eq:NLoperator}
%\end{align}
%----------------------------------------

The bistable dissipative solitons studied here are stationary solutions of Eq.~(\ref{eq:system}), and therefore we will focus on the stationary problem. To perform the bifurcation analysis of the stationary problem $\partial \mathcal{U}/\partial t=0$, we define the new variable $U=(A_R,\ A_I,\ B_R,\  B_I,\ \partial_\tau A_R,\ \partial_\tau A_I,$ $ \ \partial_\tau B_R,\ \partial_\tau B_R)$, where subscripts $R,I$ indicate the real and imaginary parts of the complex fields, respectively. Following \cite{Parra-Rivas:2021}, we  can recast the stationary problem into the dynamical system form (i.e., system of ordinary differential equations, ODEs):
\begin{equation}
    \dfrac{d U}{d \tau}=f(U;\alpha_{1,2}, \eta_{1,2}, \gamma_{ij}',\kappa, S, \sigma)\ \ \ ,
    \label{eq:ODE}
\end{equation} where the functional $f$ is defined as:
\begin{equation}
\begin{split}
f_m &= U_{m+4}, \hspace{2cm}{m=1,\cdots, 4} \\
f_5 &=\eta_1^{-1}\left[-\alpha_1U_2-\Delta U_1+\kappa\left( U_1U_3+U_2U_4\right)+\sigma \mathfrak{K}_I\right]  \\
f_6 &=\eta_1^{-1}\left[\alpha_1 U_1 -\Delta U_2 -S+\kappa\left( U_1U_4-U_2U_3\right) +\sigma \mathfrak{K}_{II} \right]  \\
f_7 &= \eta_2^{-1}\left[-\alpha_2U_4-2\Delta U_3+\kappa \left(U_1^2 - U_2^2\right) +\sigma \mathfrak{K}_{III} \right] \\
f_8 &= \eta_2^{-1}\left[\alpha_2 U_3-2\Delta U_4 + 2\kappa U_1 U_2 + \sigma \mathfrak{K}_{IV} \right] \\
\end{split}
\end{equation}

%\begin{equation}
 %   f=\begin{pmatrix}
  %      U_5\\
   %     U_6\\
    %    U_7\\
     %   U_8\\
      %  \eta_1^{-1}\left[-\alpha_1U_2-\Delta U_1+\kappa\left( U_1U_3+U_2U_4\right)+\sigma \mathfrak{K}_I\right]\\
        %\eta_1^{-1}\left[\alpha_1 U_1 -\Delta U_2 -S+\kappa\left( U_1U_4-U_2U_3\right) +\sigma \mathfrak{K}_{II} \right]\\
        %\eta_2^{-1}\left[-\alpha_2U_4-2\Delta U_3+\kappa \left(U_1^2 - U_2^2\right) +\sigma \mathfrak{K}_{III} \right]\\
        %\eta_2^{-1}\left[\alpha_2 U_3-2\Delta U_4 + 2\kappa U_1 U_2 + \sigma \mathfrak{K}_{IV} \right]
        %\end{pmatrix} \ \ \ ,
    %\label{eq:functional}
%\end{equation}
\noindent and the Kerr contribution $\mathfrak{K}$ is:
\begin{equation}
     \mathfrak{K}\equiv
        \begin{pmatrix}
            \mathfrak{K}_I\\
            \mathfrak{K}_{II}\\
            \mathfrak{K}_{III}\\
            \mathfrak{K}_{IV}\\
        \end{pmatrix} =
        \begin{pmatrix}
            \left( \gamma_1' \left( U_1^2 + U_2^2\right) +2 \gamma_{12}' \left( U_3^2+U_4^2 \right)   \right)U_1\\
            \left( \gamma_1' \left( U_1^2 + U_2^2\right) +2 \gamma_{12}' \left( U_3^2+U_4^2 \right)   \right)U_2\\
            \left( \gamma_2' \left( U_3^2 + U_4^2\right) +2 \gamma_{21}' \left( U_1^2+U_2^2 \right)   \right)U_3\\
             \left( \gamma_2' \left( U_3^2 + U_4^2\right) +2 \gamma_{21}' \left( U_1^2+U_2^2 \right)   \right)U_4\\
        \end{pmatrix}\ \ \ .
\end{equation}
\noindent 
For further details, we address the reader to Ref.~\cite{Parra-Rivas:2021}, where the particular purely quadratic case $\sigma=0$ is treated. 

Among the different classes of localized states, let us focus here on single-spike solitons, i.e., CSs. The origin of these states is commonly related to the {\it uniform-pattern bistability} (i.e., the coexistence of a stable modulated pattern with a uniform state) and the occurrence of a homoclinic bifurcation (see \cite{Parra-Rivas:2018}).

To analyze the system (\ref{eq:ODE}) and its bifurcation structure, we performed a path-continuation analysis \cite{doedel:1991} by means of the software package, i.e., AUTO-07p \cite{doedel:2009}. 
This technique allows us to study the mathematical evolution of a given solution corresponding to an ODE problem, subject to a continuous change of some dynamical parameter. As a result, bifurcations phase diagrams can be drawn, and a linear stability analysis can be performed.
%Bifurcations phase diagram can be drawn and linear stability analysis can be performed.
Starting with a CS static solution we first determine loci of the limit points of the CS existence regions. These points correspond to folds bifurcations and are computed by simultaneously tracking such point in the two-parameters ($\Delta$, $S$). Once this existence region is computed, we study the one-dimensional bifurcation diagrams associated with these states, i.e., by means of $S$ path-continuations made for different values of $\Delta$. 

The stability of these states is afterward computed through a linear stability analysis. 

From the spectrum $\boldsymbol\lambda=\{\lambda_1,..,\lambda_n\}$ of the Jacobian operator $\mathcal{J}_{ij}= \partial \mathcal{F}_i/\partial u_j$ of the system (\ref{eq:system}), where $u=(A_R,A_I,B_R, B_I)$, it is possible to retrieve the stability of the solutions found. If the real part of $\lambda_j$, $\Re\{\lambda_j\}$, is such that $\Re\{\lambda_j\}<0 \ \forall \ \lambda_j$, any small perturbation $\delta U_s$ of the considered stationary state $U_s$ tends to vanish, that is, $\delta U_s\rightarrow 0$ for $t\rightarrow\infty$. As a result, $U_s$ is stable. Furthermore, if there exists a pair of eigenvalues such that $|\Im\{\lambda_j\}|>0$ and $\Re\{\lambda_j\}>0$, the CS amplitude oscillates with a frequency which is $\propto |\Im\{\lambda_j\}|$. This is the breathing regime of the CSs. The onset of this behavior is related to a Hopf bifurcation taking place when simultaneously $\Re\{\lambda_j\}=0$ and $\Im\{\lambda_j\}=\pm i\omega$, with $\omega>0$ 

\begin{figure}[!ht]
\includegraphics[width=0.5\linewidth]{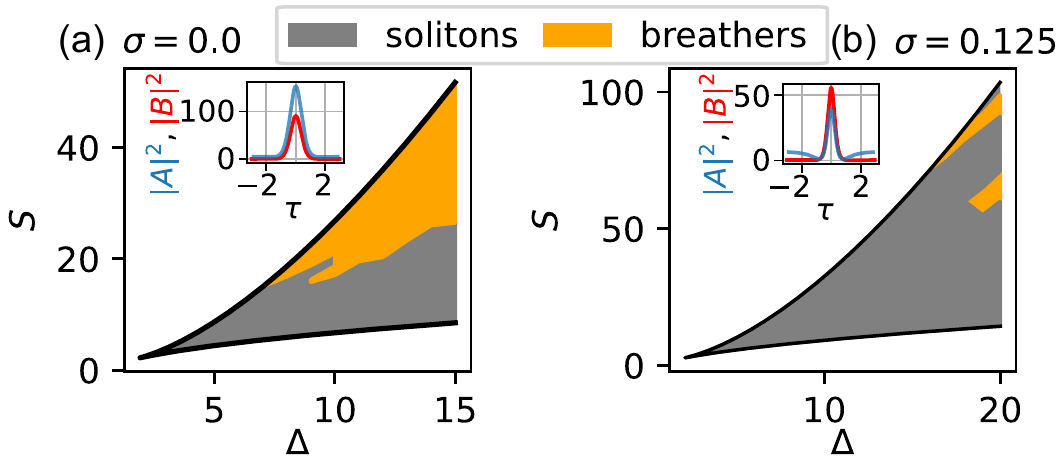}
\caption{Phase diagram in the $\left( \Delta, S\right)$ parameter space for $\sigma=0$  (a) $\sigma=0.125$  (b), respectively. In both panels, the black lines are the folds defining the region of existence of the CSs: the orange and gray colors are associated with regions of breathers or stable CSs, respectively. In the two insets, are reported two stable solutions found for $(\Delta,S)=(6,13)$. }
\label{fig:bif_000_0125}
\end{figure}

Our analysis has been performed, at first, in a purely quadratic case ($\sigma=0$). The $(\Delta,S)$-phase diagram corresponding to this case is shown in Fig.~\ref{fig:bif_000_0125} (a).  Here, the orange color denotes localized states in the breathing regime, and in gray we mark the region of stable (i.e., static) CSs. By \textit{switching on} the Kerr operator $\mathcal{K}$, i.e., by path-continuing $\sigma$ up to $\sigma=0.125$, we can compute the phase diagram of Fig.~\ref{fig:bif_000_0125} (b). 
Surprisingly, we may note that in the continuous transition $\sigma=0.0\rightarrow 0.125$ the region of CS stability broadens, thus limiting the breathing regime.

\begin{figure}[!ht]

%%%%%%%%%%%%%%%%%%%%%%%%%%%%%%%%%%%%%%%%%%%%%%%%%%%%%%%%%%%%%%%
\includegraphics[width=0.5\linewidth]{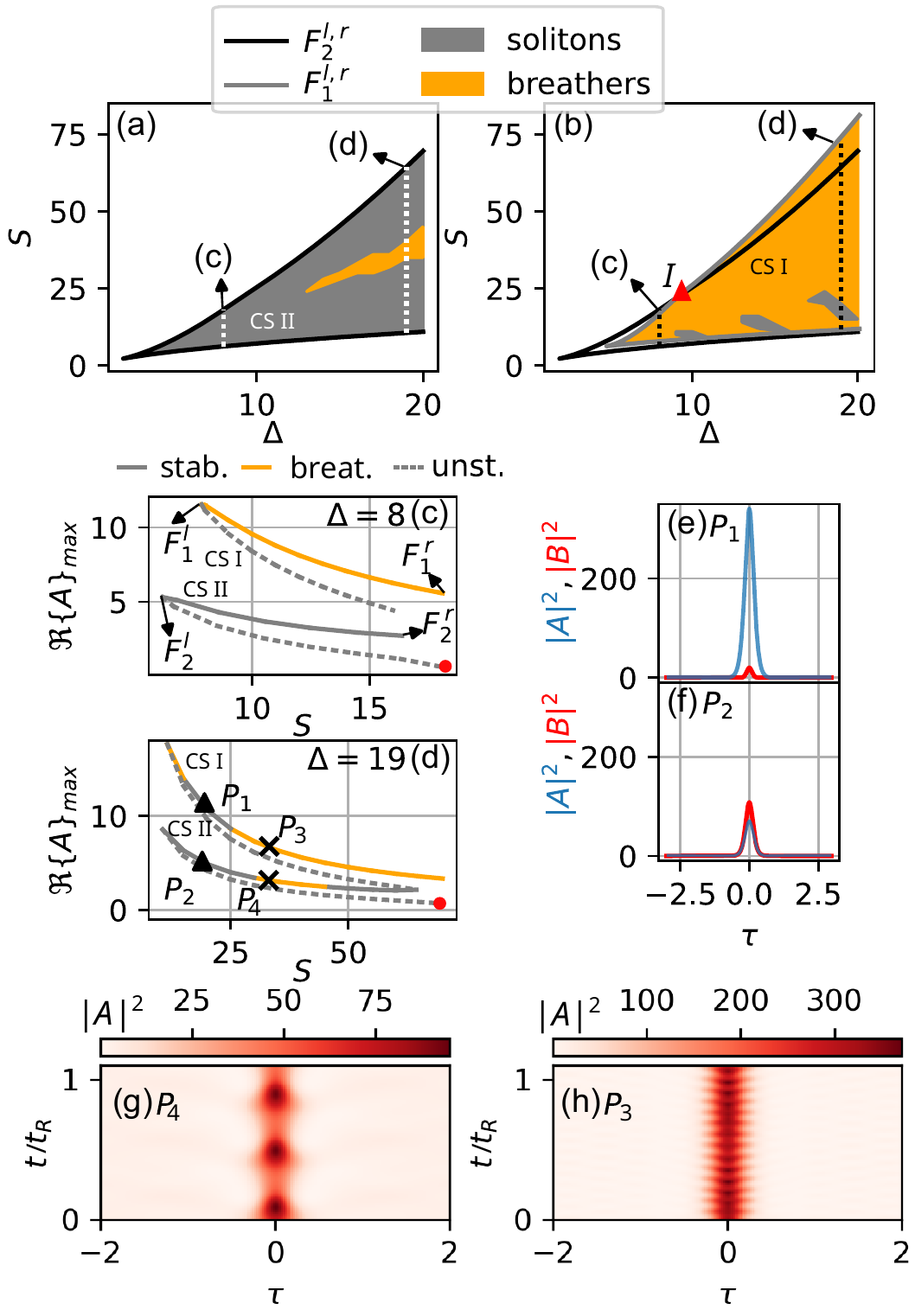}

\caption{ Bifurcations diagram of Eq.(\ref{eq:system}) for $\sigma=$ 0.25. (a) CS II and (b) CS I folds. The white and black dashed line represent  $S$ path continuations shown in panel (c,d), performed for $\Delta=8$ and $19$. (e,f) Two specific bistable CSs solutions ($P_1$ and $P_2$). (g,h) Two specific bistable breathers ($P_3$ and $P_4$). }
\label{fig:bif_025}
\end{figure}

For this value of $\sigma$, CSs bistability, like the one previously demonstrated in  Fig.~\ref{fig:DNS_pcolor}, is not observed.
In order to reach that regime, we increased the relative strength of the Kerr effect up to $\sigma=0.25$. These results are depicted in the $(\Delta, S)$-phase diagrams shown Figs.~\ref{fig:bif_025}(a) and (b). The existence regions of these states are limited by the fold bifurcations $F_2^{l,r}$ for CSs II [see Fig.~\ref{fig:bif_025}(a)] and $F_1^{l,r}$ for CSs I [see Fig.~\ref{fig:bif_025}(b)]. The overlapping of these two regions is illustrated in Fig.~\ref{fig:bif_025}(b).

%in  In this case, we discovered the existence of two classes of CS solitons (see Fig. \ref{fig:bif_025}. The associated two different folds in the parameter space $(\Delta, S)$, as shown in panel (a) (CSs I, fold 1) and (b) (CSs I , fold 2) of Fig. \ref{fig:bif_025}. 

The CSs II family results from a direct continuation of the parameter $\sigma$ from $\sigma=0$, followed by a two-parameters $(\Delta, S)$ continuation of the solutions of system $(\ref{eq:system})$. The intersection point $I\equiv(\bar\Delta, \bar S)=(8.9,21.5)$ between the two folds [see the red triangle marker on panel (b)] may be related to a high-codimension point connecting different families of CSs solutions.%, an hypothesis that must be  further confirmed. 

For $\Delta>\bar\Delta$, CSs II are connected to CSs I, and they can be located by a standard path-continuation analysis on the CS II class of solutions. This result is illustrated in the $\Re \{A\}_{max}$ (i.e. the real part of the CS peak) versus $S$ diagram depicted in 
Fig.~\ref{fig:bif_025}(d) for $\Delta=19$, which corresponds to the most right vertical white and black dashed lines in Figs.~\ref{fig:bif_025}(a),(b).  
Generally, a bifurcation diagram illustrating the modification of the field energy may be appropriate. In this specific case, however, we decided to report the $\Re \{A\}_{max}$ vs $S$ diagram, since it provides a more direct picture of the properties of the CS structures .
%Typically, a energy diagram (i.e., $L^2$-normed quantity) has a more insightful physical meaning. In this specific case, however, we decided to report $\Re \{A\}_{max}$ vs. $S$ since it illustrates better the key message we aim to communicate. 
In this diagram we have marked stable CSs with gray solid lines, breathers with orange solid curves, and unstable solutions with gray dashed lines. The connection of these two families occurs through a common fold bifurcation, yielding a multi-stable structure where unstable and stable/breathing regimes are alternating. It is worth mentioning that, for the sake of simplicity, we did not indicate the presence of a chaotic transition branch connecting the two families CSI$\rightarrow$CSII. 

Two specific bistable CSs ($P_1$ and $P_2$) are reported in Fig.~\ref{fig:bif_025}(e) and (f). For CS I [see panel (e)], most of the energy is carried by the FF wave, while for CS II [see panel (f)], the SH contribution is the highest (even if not dominant as in Fig.\ref{fig:DNS_pcolor}(f)). A similar conclusion was previously drawn (see Fig.~\ref{fig:DNS_pcolor}) by means of DNS simulations. Such type of coexisting solutions was previously predicted for competing nonlinearities in a conservative setting \cite{buryak:1995}: our study extends their findings to the case of a passive and coherently driven optical cavity. 
For a better comparison with the DNS results reported in Fig.\ref{fig:DNS_pcolor}, we should have computed the CS bifurcation structure for the same $\sigma$. However, due to the complexity of the $\chi^{(2)}+\chi^{(3)}$ system, the path continuation $\sigma=0\rightarrow1$ results numerically very challenging. Inversely, one could have performed the DNSs in Fig.~\ref{fig:DNS_pcolor} for $\sigma=0.25$. Here, however, the excitation of bistable CSs is not trivial, since the CSI and CSII energy levels are very close and one family (CSII) is a stronger dynamical attractor. 
%The bistable CSs are easily accessible, on the other side, for the case $\sigma=1$, since the resulting CSI and CSII energy levels are clearly distinguishable and further away.

Besides, we may observe bistability regions not only for static CSs but also for breathers. The dynamics of two of such states is depicted in in Figs.~\ref{fig:bif_025}(g) and (h) which correspond to the $P_3$ and $P_4$ points in Fig.~\ref{fig:bif_025}(d). We may notice that the CS I breather oscillates considerably faster and at a much higher intensity than its CS II counterpart. As a practical consequence, a perturbation of the system (e.g., by small variation of $\Delta$ or $S$) might push the breather outside its region of stability, letting the system collapse onto the CS II class of solutions.

Once the CS I are found, one is able to follow them also for $\Delta<\bar \Delta$, i.e., in a domain where CS I and CS II are no longer connected. It is thus possible that a bistable regime also exists for smaller $\sigma$ values. The situation illustrated for $\Delta=8$ in Fig.~\ref{fig:bif_025}(c), corresponds to the most left white and black dashed lines in Fig.~\ref{fig:bif_025}(a),(b). The upper branch associated with CS II is completely stable, while its CS I counterpart is unstable towards breather states. In general, the CSII group is directly connected to a homogeneous steady state (red circle marker in Fig.~\ref{fig:bif_025}(c,d)) through an unstable solution branch; we did not observe higher energy CSs families.

An interesting open question is which set of realistic parameters results in a bistable CSs state with distinguishable CSI and CSII families. $ \chi^{(2)}+\chi^{(3)}$ micro-combs have been demonstrated in Silicon Nitride \cite{Xue:2017}, Aluminium nitride \cite{lu:2023} and litihum niobate rings \cite{bruch:2021}; other interesting materials are III-V semiconductors. Our preliminary calculations for SHG-Kerr comb generation in AlGaAs microrings show that bistable CSs as in case $\sigma=1$ (Fig. \ref{fig:DNS_pcolor}) should be observable for optical losses $\alpha \sim 1/2$ dB/cm.

Optical bistability is a paradigmatic signature of nonlinear systems. Typically, it results in a low-power homogeneous state and a high power chaotic state or soliton. Eventually, multi-stable homoclinic snacking structures can arise, connecting  periodic patterns to the homogeneous solution \cite{Parra-Rivas:2018}. 
Optical bistable CSs may result from the nonlinear interaction of a FF with its harmonic waves in multi-envelope models in both single pass \cite{buryak:1995} and cavity \cite{Hansson:2023} systems. Here, we extend their existence to the case of $\chi^{(2)}+\chi^{(3)}$ systems.

\textbf{Fundings} ANR-22-CE92-0065, DFG-505515860 (Quadcomb); EU - NRRP, NextGenerationEU (PE00000001 - program “RESTART”).

\textbf{Disclosures}  The authors declare no conflicts of interest.

\bibliography{sample}

\end{document}